\documentclass[preprint,aps,prc,showpacs,nofootinbib]{revtex4}%
\usepackage{epsfig}
\usepackage{amsmath,graphicx}
\usepackage{amsmath}
\usepackage{graphicx}
\usepackage{amsfonts}
\usepackage{amssymb}
\begin{document}
\title{Rare decay $\pi^{0}\rightarrow e^{+}e^{-}$ as a Test of Standard Model }

\author{A. E. Dorokhov}
\affiliation{Joint Institute for Nuclear Research,
Bogoliubov Laboratory of Theoretical
Physics, 141980 Dubna, Moscow region, Russian Federation;\\
Institute for Theoretical Problems of Microphysics, Moscow State University, RU-119899, Moscow, Russian Federation}

\date{\today}

\begin{abstract}
Experimental and theoretical progress concerning the rare decay
$\pi^{0}\rightarrow e^{+}e^{-} $ is briefly reviewed. It includes the latest data
from KTeV and a new model independent estimate of the decay branching which show the
deviation between experiment and theory at the level of $3.3\sigma$.
The predictions for $\eta$ and $\eta'$ decays into lepton pair are presented. We also comment on the impact on
the pion rare decay estimate of the
data of BABAR collaboration on the pion transition form factor at large momentum transfer.
\end{abstract}
\maketitle

\section{Introduction}

Astrophysics observables tell us that $95\%$ of the matter in the Universe is not
described in terms of the Standard Model (SM) matter. Thus, the search for the traces of
New Physics is a fundamental problem of particle physics. There are two strategies to
look for the effects of New Physics: experiments at high energy and experiments at low
energy. In high-energy experiments it is considered that due to a huge amount of energy
the heavy degrees of freedom presumably characteristic of the SM extension sector are
possible to excite. In low-energy experiments it is huge statistics that compensates the
lack of energy by measuring the rare processes characteristic of such extensions. At
present, there is no any evidence for deviation of SM predictions from the results of
high-energy experiments and we are waiting for the LHC epoch. On the other hand, in
low-energy experiments there are rough edges indicating such deviations. The most famous
example is the muon $(g-2)$. Below it will be shown that due to recent experimental and
theoretical progress the rare process $\pi^{0}\rightarrow e^{+}e^{-} $ became a good SM
test process and that at the moment there is a discrepancy between the SM prediction and
experiment at the level of $3.3\sigma$ deviation.

\section{KTeV data}

In 2007, the KTeV collaboration published the result \cite{Abouzaid:2007md} for the
branching ratio of the pion decay into an electron-positron pair \begin{equation}
B_{\mathrm{no-rad}}^{\mathrm{KTeV}}\left(  \pi^{0}\rightarrow e^{+}e^{-}\right)
=\left(  7.48\pm0.38\right)  \cdot10^{-8}. \label{KTeV} \end{equation}
The result is based on observation of 794 candidate $\pi^0 \rightarrow e^+e^-$ events
using $K_L \rightarrow 3\pi^0$ as a source of tagged $\pi^0$s. Due to a complicated chain
of the process and a good technique for final state resolution used by KTeV this is a
process with low background.

\section{Classical theory of $\pi^{0}\rightarrow e^{+}e^{-}$ decay}

The rare decay $\pi^{0}\rightarrow e^{+}e^{-}$ has been studied theoretically
over the years, starting with the first prediction of the rate by Drell
\cite{Drell59}. Since no spinless current coupling of quarks to leptons
exists, the decay is described in the lowest order of QED as a one-loop
process via the two-photon intermediate state, as shown in Fig. 1. A factor of
$2\left(  m_{e}/m_{\pi}\right)  ^{2}$ corresponding to the approximate
helicity conservation of the interaction and two orders of $\alpha$ suppress
the decay with respect to the $\pi^{0}\rightarrow\gamma\gamma$ decay, leading
to an expected branching ratio of about $10^{-7}$. In the Standard Model
contributions from the weak interaction to this process are many orders of
magnitude smaller and can be neglected.

\begin{figure}[th]
\includegraphics[width=5cm]{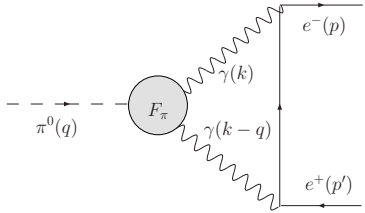}\caption{Triangle diagram for the $\pi
^{0}\rightarrow e^{+}e^{-}$ process with a pion $\pi^{0}\rightarrow
\gamma^{\ast}\gamma^{\ast}$ form factor in the vertex.}%
\label{fig:triangle}%
\end{figure}

To the lowest order in QED the normalized branching ratio is given by%
\begin{eqnarray}
R\left(  \pi^{0}\rightarrow e^{+}e^{-}\right)  =\frac{B\left(  \pi
^{0}\rightarrow e^{+}e^{-}\right)  }{B\left(  \pi^{0}\rightarrow\gamma
\gamma\right)  }=2\left(  \frac{\alpha}{\pi}\frac{m_{e}}{m_{\pi}}\right)
^{2}\beta_{e}\left(  m_{\pi}^{2}\right)  \left\vert \mathcal{A}\left(  m_{\pi
}^{2}\right)  \right\vert ^{2},
\label{B}
\end{eqnarray}
where $\beta_{e}\left(  q^{2}\right)  =\sqrt{1-4\frac{m_{e}^{2}}{q^{2}}}$,
$B\left(  \pi^{0}\rightarrow\gamma\gamma\right)  =0.988$.
The amplitude $\mathcal{A}$ can be written as%

\begin{eqnarray}
\mathcal{A}\left(  q^{2}\right)=\frac{2i}{q^{2}}\int\frac{d^{4}k}{\pi^{2}}
F_{\pi\gamma^{\ast}\gamma^{\ast}}\left(  k^{2},\left(  k-q\right)  ^{2}\right)
\frac{q^{2}k^{2}-\left(  qk\right)  ^{2}}
{\left(  k^{2}+i\varepsilon\right)
\left(  \left(  k-q\right)  ^{2}+i\varepsilon\right)  \left(  \left(
k-p\right)  ^{2}-m_{e}^{2}+i\varepsilon\right)},\\ \label{R}
\end{eqnarray}
where $q^{2}=m_{\pi}^{2},p^{2}=m_{e}^{2}$.
$F_{\pi\gamma^{\ast}\gamma^{\ast}}$ is the form
factor of the transition $\pi^{0}\rightarrow\gamma^{\ast}\gamma^{\ast}$ with
off-shell photons.

The imaginary part of $\mathcal{A}$ is defined uniquely as
\begin{eqnarray}
&&\mathrm{Im}\mathcal{A}\left(  q^{2}\right)  =\frac{\pi}{2\beta
_{e}\left(  q^{2}\right)  }\ln\left(  y_{e}\left(  q^{2}\right)  \right)
,\label{Im}\\
&& y_{e}\left(  q^{2}\right)  =\frac{1-\beta_{e}\left(  q^{2}\right)
}{1+\beta_{e}\left(  q^{2}\right)  }.\nonumber%
\end{eqnarray}
It comes from the contribution of real photons in the intermediate state and is
model independent since $F_{\pi\gamma^{\ast}\gamma^{\ast}}\left(  0,0\right)
=1$. Using inequality $\left\vert \mathcal{A}\right\vert ^{2}\geq\left(
\mathrm{Im}\mathcal{A}\right)  ^{2}$
one can get the well-known unitary bound for the branching ratio \cite{Berman60}%
\begin{eqnarray}
B\left(  \pi^{0}\rightarrow e^{+}e^{-}\right)
\geq B^{\mathrm{unitary}}\left(  \pi^{0}\rightarrow e^{+}e^{-}\right)  =4.69\cdot10^{-8}.
 \label{UnitB}\end{eqnarray}

One can attempt to reconstruct the full amplitude by using
a once-subtracted dispersion relation \cite{Bergstrom:1983ay}
\begin{equation}
\mathcal{A}\left(  q^{2}\right)  =\mathcal{A}\left(  q^{2}=0\right)
+\frac{q^{2}}{\pi}\int_{0}^{\infty}ds\frac{\mathrm{Im}\mathcal{A}\left(
s\right)  }{s\left(  s-q^{2}\right)  }.\label{DispRel}%
\end{equation}
If one assumes that Eq. (\ref{Im}) is valid for any $q^2$, then
one arrives for $q^{2}\geq4m_{e}^{2}$ at
\cite{D'Ambrosio:1986ze,Savage:1992ac,Ametller:1993we}%
\begin{eqnarray}
\mathrm{Re}\mathcal{A}\left(  q^{2}\right)=\mathcal{A}\left(
q^{2}=0\right) +\frac{1}{\beta_{e}\left(  q^{2}\right)  }
 \left[  \frac{1}%
{4}\ln^{2}\left(  y_{e}\left(  q^{2}\right)  \right)  +\frac{\pi^{2}}%
{12}+\mathrm{Li}_{2}\left(  -y_{e}\left(  q^{2}\right)  \right)  \right],
\label{Rqb} \end{eqnarray}
where $\mathrm{Li}_{2}\left(  z\right)  =-\int_{0}^{z}\left(  dt/t\right)
\ln\left(  1-t\right)  $ is the dilogarithm function.
The second term in Eq.~(\ref{Rqb}) takes into account a strong $q^{2}$
dependence of the amplitude around the point $q^{2}=0$ occurring due to the
branch cut coming from the two-photon intermediate state. In
the leading order in $\left(  m_{e}/m_{\pi}\right)  ^{2},$ Eq. (\ref{Rqb})
reduces to
\begin{equation}
\mathrm{Re}\mathcal{A}\left(  m_{\pi}^{2}\right)  =\mathcal{A}\left(
q^{2}=0\right)  +\ln^{2}\left(  \frac{m_{e}}{m_{\pi}}\right)  +\frac{\pi^{2}%
}{12}.
\end{equation}

Thus, the amplitude is fully reconstructed up to a subtraction constant.
Usually, this constant containing the nontrivial dynamics of the process is
calculated within different models describing the form factor $F_{\pi}%
(k^{2},q^{2})$
\cite{Bergstrom:1982zq,Bergstrom:1983ay,Savage:1992ac,Efimov:1981vh,Dorokhov:2007bd}.
However, it has recently been shown
in \cite{Dorokhov:2007bd} that this constant may be expressed in terms of the inverse
moment of the pion transition form factor given in symmetric kinematics of
spacelike photons%
\begin{eqnarray}
\mathcal{A}\left(  q^{2}=0\right)  =3\ln\left(  \frac{m_{e}}{\mu}\right)
-\frac{5}{4}
-\frac{3}{2}\left[  \int_{0}^{\mu^{2}}dt\frac{F_{\pi\gamma^{\ast}\gamma^{\ast
}}\left(  t,t\right)  -1}{t}
+\int_{\mu^{2}}^{\infty}dt\frac{F_{\pi\gamma
^{\ast}\gamma^{\ast}}\left(  t,t\right)  }{t}\right].\label{R0}
\end{eqnarray}
Here, $\mu$ is an arbitrary (factorization) scale. One has to note that the
logarithmic dependence of the first term on $\mu$ is compensated by the scale
dependence of the integrals in the brackets. In this way two independent processes becomes
related.

\section{Importance of CLEO data on $F_{\pi\gamma^{\ast}\gamma}$}

In order to estimate the integral in Eq.~(\ref{R0}), one needs to define
the pion transition form factor in symmetric kinematics for spacelike photon
momenta. Since it is unknown from the first principles, we will adapt the
available experimental data to perform such estimates. Let us first use the
fact that $F_{\pi\gamma^{\ast}\gamma^{\ast}}\left(  t,t\right)  <F_{\pi
\gamma^{\ast}\gamma^{\ast}}\left(  t,0\right)  $ for $t>0$ in order to obtain
the lower bound of the integral in Eq.~(\ref{R0}). For this purpose, we take
the experimental results from the CELLO \cite{Behrend:1990sr} and CLEO
\cite{Gronberg:1997fj} Collaborations for the pion transition form factor in
asymmetric kinematics for spacelike photon momentum which is well
parameterized by the monopole form \cite{Gronberg:1997fj}
\begin{eqnarray}
F_{\pi\gamma^{\ast}\gamma^{\ast}}^{\mathrm{CLEO}}\left(  t,0\right)  =\frac
{1}{1+t/s_{0}^{\mathrm{CLEO}}},\label{VMD1}\\
 s_{0}^{\mathrm{CLEO}}=\left(
776\pm22\quad\mathrm{MeV}\right)  ^{2}.\nonumber
\end{eqnarray}

For this type of the form factor one finds from Eq.~(\ref{R0})
that
\begin{eqnarray}
\mathcal{A}\left(  q^{2}=0\right)
 >-\frac{3}{2}\ln\left(  \frac
{s_{0}^{\mathrm{CLEO}}}{m_{e}^{2}}\right)  -\frac{5}{4}=-23.2\pm
0.1.\label{Rcleo}
\end{eqnarray}
Thus, for the branching ratio we are able to establish the important lower
bound which considerably improves the unitary bound given by Eq.~(\ref{UnitB})
\begin{eqnarray}
B\left(  \pi^{0}\rightarrow e^{+}e^{-}\right)
 >B^{\mathrm{CLEO}}\left(
\pi^{0}\rightarrow e^{+}e^{-}\right)  =\left(  5.84\pm0.02\right)\cdot10^{-8}.
\label{Bcleo}\end{eqnarray}

It is natural to assume that the monopole form is
also a good parametrization for the form factor in symmetric kinematics
\begin{eqnarray}
F_{\pi\gamma^{\ast}\gamma^{\ast}}\left(  t,t\right)  =\frac{1}{1+t/s_{1}}.\label{Ftt}%
\end{eqnarray}
The scale $s_{1}$ can be fixed from the relation for the slopes of the form
factors in symmetric and asymmetric kinematics at low $t$
\cite{Dorokhov:2003sc},
\begin{equation}
\left.  -\frac{\partial F_{\pi\gamma^{\ast}\gamma^{\ast}}\left(  t,t\right)
}{\partial t}\right\vert _{t=0}=\left.  -2\frac{\partial F_{\pi\gamma^{\ast
}\gamma^{\ast}}\left(  t,0\right)  }{\partial t}\right\vert _{t=0}%
,\label{slope}%
\end{equation}
that gives $s_{1}=s_{0}/2$. Note that a similar reduction of the scale is also
predicted by OPE QCD from the large momentum behavior of the form factors:
$s_{1}^{OPE}=s_{0}^{OPE}/3$ \cite{Lepage:1980fj}. Thus, the estimate for
$\mathcal{A}\left(  0\right)$ can be obtained from
Eq.~(\ref{Rcleo}) by shifting the lower bound by a positive number which
belongs to the interval $[3\ln(2)/2,3\ln(3)/2]$
\begin{equation}
\mathcal{A}\left(  q^{2}=0\right)  =-\frac{3}{2}\ln\left(  \frac{s_{1}}%
{m_{e}^{2}}\right)  -\frac{5}{4}=-21.9\pm0.3.\label{R0t}%
\end{equation}
With this result the branching ratio becomes
\begin{equation}
B\left(  \pi^{0}\rightarrow e^{+}e^{-}\right)  =\left(  6.23\pm0.09\right)
\cdot10^{-8}.\label{Bt}%
\end{equation}
This is $3.3$ standard deviations lower than the KTeV result given by
Eq.~(\ref{KTeV}).

\section{Other decay modes}
The $\eta\rightarrow l^{+}l^{-}$ decay can be analyzed in a similar manner.
As in the pion case, the CLEO Collaboration has parameterized the data for the
$\eta$-meson in the monopole form \cite{Gronberg:1997fj}:
\begin{eqnarray}
F_{\eta\gamma^{\ast}\gamma^{\ast}}^{\mathrm{CLEO}}\left(  t,0\right)
=\frac{1}{1+t/s_{0\eta}^{\mathrm{CLEO}}},\\
 s_{0\eta}^{\mathrm{CLEO}}=\left(  774\pm29\quad\mathrm{MeV}\right)  ^{2},
\nonumber\end{eqnarray}
which is very close to the relevant pion parameter. Then following the previous
case (with evident substitutions), one finds the bounds for the $q^{2}%
\rightarrow0$ limit of the amplitude $\eta\rightarrow\mu^{+}\mu^{-}$ as
\begin{eqnarray}
\mathcal{A}_{\eta}\left(  q^{2}=0\right)
>-\frac{3}{2}\ln\left(
\frac{s_{0\eta}^{\mathrm{CLEO}}}{m_{\mu}^{2}}\right)  -\frac{5}{4}=-\left(
7.2\pm0.1\right)  ,
\end{eqnarray}
and for $\eta\rightarrow e^{+}e^{-}$ one gets again Eq.~(\ref{Rcleo}). The
obtained estimates allow one to find the bounds for the branching ratios
\begin{eqnarray}
B\left(  \eta\rightarrow\mu^{+}\mu^{-}\right)     <\left(  6.23\pm
0.12\right)  \cdot10^{-6},\label{RbEta}\\
B\left(  \eta\rightarrow e^{+}e^{-}\right)     >\left(  4.33\pm0.02\right)
\cdot10^{-9}.\nonumber
\end{eqnarray}
It is important to note that for the decay $\eta\rightarrow\mu^{+}\mu^{-}$
we get the upper limit
for the branching. This is because the real part of the amplitude for this
process taken at the physical point $q^{2}=m_{\eta}^{2}$ for the parameter
$s_{0\eta}^{\mathrm{CLEO}}$ remains negative and a positive shift due to the
change of the scale $s_{0\eta}\rightarrow s_{1\eta}$ reduces the absolute
value of the real part of the amplitude $\left\vert \mathrm{Re}%
\mathcal{A}\left(  m_{\eta}^{2}\right)  \right\vert $. At the same time,
considering the decays of $\pi^{0}$ and $\eta$ into an electron-positron pair,
the evolution to physical point (\ref{Rqb}) makes the real part of the
amplitude to be positive for the parameter $s_{0}^{\mathrm{CLEO}}$ and the
absolute value of the real part of the amplitude increases in changing the
scales of the meson form factors.

The predicted branchings are given in the Table and
compared with existing experimental data. The so-called unitary bound appears
if in (\ref{B}) only the imaginary part of the amplitude which is model
independent is taken into account. The CLEO bound corresponds to the estimate
of the real part of the amplitude basing on the CELLO and CLEO data on the
meson transition from factors \cite{Dorokhov:2007bd}. The fourth column of the Table contains the
predictions where in addition the constraint from OPE QCD on the transition
form factor for arbitrary photon virtualities is taken into
account\cite{Dorokhov:2007bd}.

The results that take into account the mass corrections to the amplitude \cite{Dorokhov:2009xs}, mainly due to powers $(M/\Lambda)^2$, where $M$ is the pseudoscalar meson mass and $\Lambda\approx M_\rho$ is characteristic scale of the pion transition form factor,
are given in the fifth column of the Table. They are essentially visible for $\eta\left(  \eta^{\prime
}\right)  $ meson decays. It is interesting that for $\eta$ decay to muons the
mass correction shifts the theoretical prediction in the direction to the
unitary bound and thus opposite to the experimental result \cite{Abegg:1994wx}.
Thus, it would be very desirable to check
experimentally the predicted bounds for the process
$\eta\rightarrow\mu^{+}\mu^{-}$.
  Also note the recent measurement by the WASA/Celcius Collaboration
\cite{Berlowski:2008zz} which improves
the upper limit for the branching $\eta\rightarrow e^{+}e^{-}$.

For $\eta^{\prime}$ decays there appear new thresholds in addition to
the two-photon one. In general, this violates the unitary bound because the
correction $A_z$ gains an imaginary part at $z=(M/\Lambda)^2>1$
\begin{equation}
\Delta\Im\mathcal{A}   =-\frac{\pi}{\beta}\left(  1-\frac{1}{z}\right)
^{2}\ln\left(  \frac{1+\beta}{1-\beta}\right) \Theta(z-1) .
\end{equation}
 As it is
seen from the table, this happens for the $\eta^{\prime}\rightarrow\mu\mu$
channel. Nevertheless, it turns out that the predictions for this channel are
quite accurate.

\begin{table}[th]
\caption[Results]{Values of the branchings $B\left(  P\rightarrow l^{+}%
l^{-}\right)  $ obtained in our approach and compared with the available
experimental results. }%
\label{table2}
\begin{tabular}
[c]{|c|c|c|c|c|c|}\hline
$R_{0}$ & Unitary bound & CLEO bound & CLEO+OPE \cite{Dorokhov:2007bd} & \cite{Dorokhov:2009xs} & Experiment\\
&  &  &  &  & \\\hline
$R_{0}\left(  \pi^{0}\rightarrow e^{+}e^{-}\right)  \times10^{8}$ & $\geq4.69$
& $\geq5.85\pm0.03$ & $6.23\pm0.12$ & $6.26$ & $7.49\pm0.38$
\cite{Abouzaid:2007md}\\\hline
$R_{0}\left(  \eta\rightarrow\mu^{+}\mu^{-}\right)  \times10^{6}$ & $\geq4.36$
& $\leq6.23\pm0.12$ & $5.12\pm0.27$ & $4.64$ & $5.8\pm0.8$
\cite{Abegg:1994wx}\\\hline
$R_{0}\left(  \eta\rightarrow e^{+}e^{-}\right)  \times10^{9}$ & $\geq1.78$ &
$\geq4.33\pm0.02$ & $4.60\pm0.09$ & $5.24$ & $\leq2.7\cdot10^{4}$
\cite{Berlowski:2008zz}\\\hline
$R_{0}\left(  \eta^{\prime}\rightarrow\mu^{+}\mu^{-}\right)  \times10^{7}$ &
$\geq1.35$ & $\leq1.44\pm0.01$ & $1.364\pm0.010$ & $1.30$ & \\\hline
$R_{0}\left(  \eta^{\prime}\rightarrow e^{+}e^{-}\right)  \times10^{10}$ &
$\geq0.36$ & $\geq1.121\pm0.004$ & $1.178\pm0.014$ & $1.86$ & \\\hline
\end{tabular}
\end{table}

\section{Possible explanations of the effect}

Therefore, it is
extremely important to trace possible sources of the discrepancy between the
KTeV experiment and theory. There are a few possibilities: (1) problems with
(statistic) experiment procession, (2) inclusion of QED radiation corrections
by KTeV is wrong, (3) unaccounted mass corrections are important, and (4)
effects of new physics. At the moment, the last possibilities were
reinvestigated. In \cite{Dorokhov:2008qn}, the contribution of QED radiative
corrections to the $\pi^{0}\rightarrow e^{+}e^{-}$ decay, which must be taken
into account when comparing the theoretical prediction (\ref{Bt}) with the
experimental result (\ref{KTeV}), was revised. Comparing with earlier
calculations \cite{Bergstrom:1982wk}, the main progress is in the detailed
consideration of the $\gamma^{\ast}\gamma^{\ast}\rightarrow e^{+}e^{-}$
subprocess and revealing of dynamics of large and small distances.
Occasionally, this number agrees well with the earlier prediction based on
calculations \cite{Bergstrom:1982wk} and, thus, the KTeV analysis of radiative
corrections is confirmed. In \cite{Dorokhov:2008cd,Dorokhov:2009xs} it was shown that the mass
corrections are under control and do not resolve the problem. So our main
conclusion is that the inclusion of radiative and mass corrections is unable
to reduce the discrepancy between the theoretical prediction for the decay
rate (\ref{Bt}) and experimental result (\ref{KTeV}).

\section{$\pi^{0}\rightarrow e^{+}e^{-}$ decay as a filtering process for low mass dark matter}

If one thinks about an extension of the Standard Model in terms of heavy, of an order of
$100$ GeV or higher, particles, then the contribution of this sort of particles to the
pion decay is negligible. However, there is a class of models for description of Dark
Matter with a mass spectrum of particles of an order of 10 MeV \cite{Boehm:2003hm}.
This model postulates a neutral scalar dark matter particle $\chi$
which annihilates to produce electron/positron pairs: $\chi\chi \to e^+e^-$. The excess
positrons produced in this annihilation reaction could be responsible for the bright 511
keV line emanating from the center of the galaxy \cite{Weidenspointner:2006nua}. The
effects of low mass vector boson $U$ appearing in such model of dark matter were considered in \cite{Kahn:2007ru} where the excess of KTeV data over theory
put the constraint on coupling which is consistent with that coming from the muon
anomalous magnetic moment and relic radiation \cite{FayetLatest}. Thus, the pion decay
might be a filtering process for light dark matter particles.

\section{Constraints from BABAR data}

Let us now consider the amplitude $\pi^0\to e^+e^-$ in the
context of the constituent quark model \cite{Quark1,Quark2}. Within this model,
the pion form factor is given by the quark-loop (triangle) (Fig. 1)
diagram with momentum independent constituent quark mass. The result for the form factor in
symmetric kinematics is given by \cite{Quark2}
\begin{equation}
F_{\pi\gamma^*\gamma^*}(t,t)=\frac{2M_Q^2}{\beta_Qt}\ln\frac{\beta_Q+1}{\beta_Q-1},
\end{equation}
where $\beta_Q=\sqrt{1+\frac{4M_Q^2}{t}}$,
and the amplitude at zero is
\begin{equation}
\mathcal{A}\left(  0\right)=\frac{3}{2}\ln(\frac{m_e^2}{M_Q^2})-\frac{17}{4}.
\end{equation}
It is interesting to remind that in asymmetric kinematics the form factor has double logarithmic asymptotics at large momentum transfer $log^2(Q^2/M_q^2)$  \cite{Quark2}
\begin{equation}
F_{\pi\gamma\gamma^*}(t,0)=\frac{m_\pi^2}{m_\pi^2+t}\frac{1}{2\arcsin^2(\frac{m_\pi}{2M_Q})}
\{2\arcsin^2(\frac{m_\pi}{2M_Q})+\frac{1}{2}\ln^2\frac{\beta_Q+1}{\beta_Q-1}\}.
\label{Ftt}\end{equation}

Recently BABAR collaboration announced new data \cite{photon09} on pion transition form factor $F_{\pi\gamma\gamma^*}(t,0)$ at very large $t$ up to 40 GeV$^2$. The data are in strong contradiction with massless QCD predicting asymptotic $1/t$ behavior with single logariphmic modulation in leading order. However,
the BABAR data are well fitted by (\ref{Ftt}) if the parameter $M_Q$ is taken as $M_Q=135$ MeV. Then the amplitude at zero is
$\mathcal{A}\left(  0\right)=-20.98$ which is still far from numbers corresponding to KTeV: $\mathcal{A}\left(  0\right)=-18.6\pm0.9$. In \cite{Pivovarov:2001mw} it was also noted that in order to get reasonable estimate of the hadronic vacuum polarization to the anomalous magnetic moment of muon it is necessary to take the quark mass rather small $M_Q=180$ MeV, the value which is close to the pion mass.

Within perturbative QCD one possible scenario to explain the form factor growth faster than expected is to assume that the full resummation of perturbative logarithms is important\footnote{I am indebted to A.A. Pivovarov for discussion this point.}. Indeed, sometime ago Manohar augmented \cite{Manohar:1990hu} that for the pion transition form factor in kinematics when one photon is virtual and other is quasireal the resummation effects dominate its behavior at large fixed $Q^2$ and probably BABAR (if confirmed) observes these effects (logarithmic or powerlike).

5\label{fig:largenenough}

\begin{figure}[th]
\includegraphics[width=9cm]{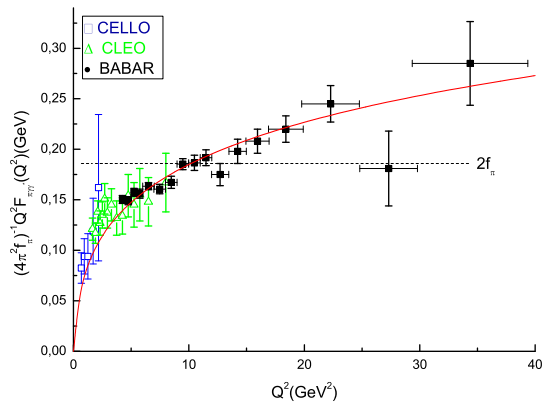}\caption{The transition form factor $\pi
^{0}\rightarrow \gamma^\ast\gamma$. The data are from CELLO \cite{Behrend:1990sr},
CLEO \cite{Gronberg:1997fj} and BABAR \cite{photon09} Collaborations.
The dashed line is massless QCD asymptotic limit. The solid line is the form factor calculated from massive triangle
diagram, Eq. (\ref{Ftt}), with mass parameter taken as $M_q=135$ MeV.}
\label{babar09}%
\end{figure}

Further independent
experiments at KLOE, NA48, WASAatCOSY, BES III and other facilities will be crucial for
resolution of the problem with the rare leptonic decays of light pseudoscalr mesons. Also it is important to confirm the theoretical base for maximally model independent prediction of the branchings by getting more precise data on the pion
transition form factor in asymmetric as well in symmetric kinematics in wider region of momentum transfer that is soon expected from BABAR and BELLE collaborations.

\section{Acknowledgments}

We are grateful to the Organizers and personally professor V. Skalozub for a nice meeting and kind invitation
to present our results.

\end{document}